\documentclass[12pt]{article}
\newcommand{\be}{\begin{equation}}
\newcommand{\ee}{\end{equation}}
\newcommand{\beq}{\begin{equation}}
\newcommand{\eeq}{\end{equation}}
\newcommand{\ba}{\begin{eqnarray}}
\newcommand{\ea}{\end{eqnarray}}

\begin{document}
\baselineskip=15.5pt
\pagestyle{plain}
\setcounter{page}{1}


\def\del{{\partial}}
\def\vev#1{\left\langle #1 \right\rangle}
\def\cn{{\cal N}}
\def\co{{\cal O}}
\newfont{\Bbb}{msbm10 scaled 1200}     
\newcommand{\mathbb}[1]{\mbox{\Bbb #1}}
\def\IC{{\mathbb C}}
\def\IR{{\mathbb R}}
\def\IZ{{\mathbb Z}}
\def\RP{{\bf RP}}
\def\CP{{\bf CP}}
\def\Poincare{{Poincar\'e }}
\def\tr{{\rm tr}}
\def\tp{{\tilde \Phi}}

\def\TL{\hfil$\displaystyle{##}$}
\def\TR{$\displaystyle{{}##}$\hfil}
\def\TC{\hfil$\displaystyle{##}$\hfil}
\def\TT{\hbox{##}}
\def\HLINE{\noalign{\vskip1\jot}\hline\noalign{\vskip1\jot}} 
\def\seqalign#1#2{\vcenter{\openup1\jot
  \halign{\strut #1\cr #2 \cr}}}
\def\lbldef#1#2{\expandafter\gdef\csname #1\endcsname {#2}}
\def\eqn#1#2{\lbldef{#1}{(\ref{#1})}%
\begin{equation} #2 \label{#1} \end{equation}}
\def\eqalign#1{\vcenter{\openup1\jot
    \halign{\strut\span\TL & \span\TR\cr #1 \cr
   }}}
\def\eno#1{(\ref{#1})}
\def\href#1#2{#2}
\def\half{{1 \over 2}}

\def\ads{{\it AdS}}
\def\adsp{{\it AdS}$_{p+2}$}
\def\cft{{\it CFT}}

\newcommand{\ber}{\begin{eqnarray}}
\newcommand{\eer}{\end{eqnarray}}


\newcommand{\nonu}{\nonumber}
\newcommand{\oh}{\displaystyle{\frac{1}{2}}}
\newcommand{\dsl}
  {\kern.06em\hbox{\raise.15ex\hbox{$/$}\kern-.56em\hbox{$\partial$}}}
\newcommand{\id}{i\!\!\not\!\partial}
\newcommand{\as}{\not\!\! A}
\newcommand{\ps}{\not\! p}
\newcommand{\ks}{\not\! k}
\newcommand{\D}{{\cal{D}}}
\newcommand{\dv}{d^2x}
\newcommand{\Z}{{\cal Z}}
\newcommand{\N}{{\cal N}}
\newcommand{\Dsl}{\not\!\! D}
\newcommand{\Bsl}{\not\!\! B}
\newcommand{\Psl}{\not\!\! P}
\newcommand{\eeqarr}{\end{eqnarray}}
\newcommand{\ZZ}{{\rm \kern 0.275em Z \kern -0.92em Z}\;}


\begin{titlepage}

\rightline{NORDITA-2003-55 HE} 
 
\vskip 2.5cm \centerline{\Huge Gaugino condensate and phases of} 
\vskip 0.5cm 
 \centerline{\Huge  ${\cal N}=1$ super Yang-Mills theories} 
 
\vskip 1.5cm

\centerline{\large P. Merlatti} 
 
\vskip .8cm 
 
\centerline{\sl NORDITA} 
 
\centerline{Blegdamsvej 17, 2100 Copenhagen \O, Denmark} 
 
\centerline{\tt merlatti@nbi.dk} 
 
\vskip 3.5cm 
 
\begin{abstract} 
\noindent I consider ${\cal N}=1$ $U(N)$ gauge theory with matter in the adjoint, fundamental and anti-fundamental representations. Focusing on the equations defining the Riemann surface that describes the quantum theory, the gaugino condensates (and related superpotentials) are calculated in the limit of $SU(N)$ gauge group, both in the pure theory and in the presence of matter. In the case without fundamental matter it is investigated the structure of the space of vacua. In particular it is discussed how different vacua can be related, in a way which finally helps to count them.

\end{abstract}
.

\end{titlepage}

\tableofcontents
\section{Introduction}

After the work of Dijkgraaf and Vafa \cite{Dijkgraaf:2002dh}, where an interesting and useful relation between supersymmetric gauge theories and matrix models has been proposed, the dynamics of ${\cal N}=1$ supersymmetric gauge theory has been much investigated. A lot of effort has indeed been spent to test, generalize and prove \cite{Dijkgraaf:2002xd,Aganagic:2003xq,Cachazo:2002ry} the conjecture made in \cite{Dijkgraaf:2002dh}. This has brought a much deeper comprehension of the quantum dynamics of gauge theories. In particular, the study of the various physical quantities as a function of the parameters that define the theory has shown that at the quantum level there is a rich and highly non trivial structure of this parameter space \cite{Cachazo:2002zk,Cachazo:2003yc,Ferrari:2002kq,Ferrari:2003yr}. 

One of the most interesting problem of the non perturbative dynamics of such a theory remains perhaps that of gaugino condensation. In this paper we try to find results about gaugino condensation using the recently developed techniques we are talking about and we study some properties of the parameter space.

We start then by considering ${\cal N}=1$ supersymmetric $U(N)$ gauge theory with a chiral superfield $\Phi$ in the adjoint representation of the gauge group, $N_f$ matter fields $Q_f$ in the fundamental representation and $N_f$ ($\tilde{Q}_{\tilde{f}}$) in the anti-fundamental one. The superpotential is taken to be:
\begin{equation}
\label{tree}
W_{tree}(\Phi) = \frac{1}{2}M~\mbox{Tr}~\Phi^2+\tilde{Q}_{\tilde{f}}m_f^{\tilde{f}}Q^f.
\end{equation}
If we consider values of the parameters such that $0<<m_i<<M$ we can explore the theory at different and well defined ranges of energy. Indeed at energies $m_i<E<M$ the theory looks like ${\cal N}=1$ supersymmetric QCD with $N_f$ massive flavors. At energy $E<m_i$ we deal instead with pure ${\cal N}=1$ gluodynamics.

Focusing on the chiral ring, this theory has been (at least implicitly) solved in \cite{Cachazo:2003yc}. This has been possible thanks to the definition of some operators in function of which it is possible to write all the observable of the theory \cite{Cachazo:2002ry}. It has been found that these operators have to solve some equations coming from the generalized Konishi anomaly. These equations, that can be interpreted as Ward identities, arise purely at the quantum level and encode all the quantum dynamics of the theory. To explicitly solve these equations thus corresponds to determine all the observables of the theory, taking into account the full quantum dynamics.

In the first part of this paper we will deal with these equations and their implicit solution (given in \cite{Cachazo:2003yc}) for the theory we just defined. We think this is worthwhile to do, seen the interesting limits this theory has.
The main result we will get is that the quantum non-perturbative dynamic generates a condensate of the gaugino bilinear (for every $N_f<N$\footnote{For $N_f\geq N$ baryons should be included}). We will determine it. The result is in perfect agreement with previous computations, in particular with the so called weak coupling computation (see \cite{Dorey:2002ik} for an exhaustive review about this controversial computation). The computation will be made without introducing a superpotential for this superfield ($S=\mbox{Tr}(W_{\alpha}W^{\alpha})$) that would imply to treat it as an elementary field. However, to better interpret the results we find in the different limits we consider, we will use the relation of this gaugino condensate with the generation of a superpotential $W_{low}(\Lambda)$. Once we have it, we can write different effective superpotentials integrating in different fields. This procedure is now well understood (see for example \cite{Intriligator:1995au}) and its validity (in the form we will use) founds on the Intrilligator-Leigh-Seiberg (ILS) linearity principle \cite{Intriligator:1994jr}. In the case at hand the validity of this principle is further supported by the constraints coming from the global symmetries of this theory \cite{Intriligator:1995au}. We can then freely integrate in and out different fields and consider different ranges of energies depending on what we are interested in.
In this way, we will be able to compare the results with the literature. We will then derive the Veneziano-Yankielowicz superpotential for pure ${\cal N}=1$ gluodynamics and both the Veneziano-Yankielowicz-Taylor and Affleck-Dine-Seiberg superpotentials for ${\cal N}=1$ supersymmetric QCD.
However, let me stress once more that this extrapolation `off-shell' of the results is not necessary and that, at a basic level, what we find is the value of the gaugino condensate of these theories, that has not necessarily to be explained by the minimization of a superpotential for the field $S$.

In the second part of the paper we turn to consider more general patterns of breaking, obtained by means of more general choices of $W(\Phi)$
\begin{equation}
W(\Phi)~=~\sum_{k=0}^n\frac{g_k}{k+1}\mbox{Tr}~\Phi^{k+1}
\end{equation}
such that the low energy dynamics is that of a gauge theory with gauge group $\prod_{i=1}^nU(N_i)$. In this case, a superpotential for the $S_i$ cannot be derived by the ``integrating-in'' procedure as we will do for the particular choice (\ref{tree}). The reason is that, when the gauge group factorizes, the different condensates do not couple to independent sources (in the microscopic Lagrangian it appears only one gauge coupling, the one of the underlying $U(N)$ theory). However, following the lines of \cite{Cachazo:2002zk}, we will try to say something about the structure of the parameter space of such a theory. For this analysis we will concentrate on the case with $N_f=0$, so that precise order parameters for the underlying $U(N)$ theory can be well defined \cite{Cachazo:2002zk}.
We will find that, given a partition of $N_i$ and a particular vacuum where the confinement index, (as it has been defined in \cite{Cachazo:2002zk}), is $t$, it is easy to build up $m$ vacua with $t$ as confinement index, where $m$ is the minimum commune multiple of the $N_i$. We will define then a low energy discrete chiral `effective' symmetry under which these vacua are naturally permuted. This symmetry will relate different branches of the moduli space. A non trivial consequence that follows from this is that there are always a multiple of $m$ vacua for every value of the confinement index. We will finally investigate the relation of this structure with the multiplication map defined in \cite{Cachazo:2001jy}.

\section{Computation of the gaugino condensate}

We consider ${\cal N}=1$ supersymmetric $U(N)$ gauge theory with a chiral superfield $\Phi$ in the adjoint representation of the gauge group, $N_f$ matter fields $Q_f$ in the fundamental representation and $N_f$ anti-fundamental fields $\tilde{Q}_{\tilde{f}}$. The superpotential is taken to be:
\begin{equation}
\label{ourpot}
W_{tree}(\Phi) = \frac{1}{2}M~\mbox{Tr}~\Phi^2+\tilde{Q}_{\tilde{f}}m_f^{\tilde{f}}Q^f
\end{equation}
where we choose the matrix $m_f^{\tilde{f}}=~m_f\delta_f^{\tilde{f}}$ to be diagonal and such that $m_f$ are canonically normalized masses. Moreover we define $B=\mbox{Det}(m)$.

As already discussed in the introduction, in \cite{Cachazo:2003yc} it has been shown that the full quantum theory can be described by a Riemann surface and some functions (or differential forms) on it. The general solution reads (see equation (5.17) of \cite{Cachazo:2003yc}):
\begin{eqnarray}
\label{csw}
P^2(z)-\alpha B=F(z) H^2(z)\\
W'(z)^2+f=M^2F(z)
\nonumber
\end{eqnarray}
where $P(z)$ is a degree $N$ polynomial, $f$ in our case is a constant, related to the gaugino condensate $S$ by \cite{Cachazo:2001jy}:
\begin{equation}
S~=~-\frac{1}{4M}~f
\label{condf}
\end{equation}
and $\alpha/4~=~\Lambda_I^{2N-N_f}$ and $\Lambda_I$ is the dynamically generated scale of the theory, in the subtraction scheme related to the standard ${\cal N}=2$ parent theory when the interaction between the fundamental matter and the adjoint chiral superfield is neglected. Let me define also other scales related to this one by the following renormalization group matching conditions:
\begin{eqnarray}
\Lambda_{II}^{3N-N_f}~=~\Lambda_I^{2N-N_f}M^N\label{scale}\\
\Lambda_{III}^{3N}~=~\Lambda_{II}^{3N-N_f}\prod_{i=1}^{N_f}m_i
\label{scales}
\end{eqnarray}
These scales are the physical scales of the theory, each one corresponding to a different range of energy. We fix our parameters such that $0<<m_i<<M$, where all the $m_i$ ($i=1\ldots N_f$) are of the same order of magnitude. We can then study the theory in three different regimes:
\begin{itemize}
\item at energy $E>M$ the theory is the UV theory we just described, with its natural scale $\Lambda_I$
\item at energy $m_i<E<<M$ the adjoint superfield $\Phi$ decouples and we are left with ${\cal N}=1$ supersymmetric QCD with $N_f$ massive flavors. The dynamical generated scale of this theory is $\Lambda_{II}$
\item at energy $E<<m_i$ also the fundamental and anti-fundamental fields decouple and we deal then in all respects with pure ${\cal N}=1$ gluodynamics, characterized by its scale $\Lambda_{III}$.
\end{itemize}
Let's turn now to the solution of our model. Starting from the microscopic theory with superpotential (\ref{ourpot}), we can easily get from (\ref{csw}):
\begin{eqnarray}
\label{fact}P_N^2(z)-4 \Lambda_I^{2N-N_f}m^{N_f}~=~F_2(z)H_{N-1}^2(z)\\
F_2(z)~=~z^2+\frac{f}{M^2}\nonumber
\end{eqnarray}
We know \cite{Cachazo:2003yc} that solving (\ref{fact}) corresponds to solve the full theory. 

To this aim, we consider now Chebyshev polynomial (see Appendix A), in terms of which a solution to (\ref{fact}) can be found and it is:
\begin{equation}
P_N(z)~=~2\tilde{\Lambda}^N\eta^N{\cal T}_N\left(\frac{z}{2\eta\tilde{\Lambda}}\right),\hspace{0.5cm}F_2(z)~=~z^2-4\eta^2\tilde{\Lambda}^2,\hspace{0.5cm}H_{N-1}(z)~=~\eta^{N-1}\tilde{\Lambda}^{N-1}{\cal U}_{N-1}\left(\frac{z}{2\eta\tilde{\Lambda}}\right)
\end{equation}
where 
\begin{equation}
\tilde{\Lambda}^{2N}~=~\Lambda_I^{2N-N_f}\prod_{i=1}^{N_f}m_i\hspace{2cm}\mbox{and}\hspace{2cm}\eta^{2N}=1;
\end{equation} 
${\cal T}_N(z)$ and ${\cal U}_{N-1}(z)$ are the Chebyshev polynomials respectively of the first and second kind (see Appendix A). What is most important for us is that the solution also implies: 
\begin{equation}
f~=~-4\eta^2 M^2 (\Lambda_I^{2N-N_f}\prod_{i=1}^{N_f}m_i)^{1/N},
\label{f}
\end{equation}

Moreover, we know that the equation (\ref{fact}) has been proved to have a unique solution in a very analogous case \cite{Cachazo:2002pr} and that proof is also valid for the case we are considering here. From (\ref{condf}) and (\ref{f}) it is now easy to see that the gaugino condensate is non vanishing and it is equal to
\begin{equation}
S~=~\eta^2 M (\Lambda_I^{2N-N_f}\prod_{i=1}^{N_f}m_i)^{1/N}
\label{condtot}
\end{equation}

We begin to notice that for $E<<m_i$, writing thus (\ref{condtot}) in function of $\Lambda_{III}$, we get:
\begin{equation}
S~=~\eta^2\Lambda_{III}^3
\end{equation}
This is in perfect agreement with the expectations of the so called `weak coupling computation' for pure ${\cal N}=1$ gluodynamics (see \cite{Konishi:2003ts} for a related way of getting the same result; see instead \cite{Dorey:2002ik} for a general discussion).

In the next subsection, using standard techniques to integrate in and out various fields, we will relate these results to the known superpotentials.

\subsection{Determination of the superpotentials}

Consider that the field $S$ is canonically conjugated to $\ln\Lambda_I^{2N-N_f}$ and from
\begin{equation}
\frac{\partial W_{low}}{\partial \ln\Lambda_I^{2N-N_f}}~=~\langle S\rangle
\end{equation}
together with (\ref{condtot}) we can find the low energy dynamically generated superpotential
\begin{equation}
\label{pottot}
W_{low}~=~N\eta^2M~ \left(\prod_{f=1}^{N_f} m_f\right)^{1/N}\Lambda_I^{\frac{2N-N_f}{N}}
\end{equation}
We are interested in studying the $SU(N)$ dynamics of supersymmetric QCD (or eventually its low energy limit), so we will always make the limit $M\to\infty$ to decouple the adjoint chiral 
superfield. Using then (\ref{scale}) equation (\ref{pottot}) becomes:
\begin{equation}
W_{low}~=~ N\eta^2\left(\prod_{f=1}^{N_f} m_f\right)^{1/N}\Lambda_{II}^{\frac{3N-N_f}{N}}
\label{pottotads}
\end{equation}

Now that we have found that a dynamical potential is generated in supersymmetric QCD and we have derived it, we are left with different possibilities. Indeed, as already anticipated, using the ILS linearity principle \cite{Intriligator:1994jr}, whose validity in this case is supported by the constraints coming from the symmetries of the theory under discussion, we can integrate in different fields. According to which fields we integrate in and to which range of energy we consider, we will get different superpotentials.

Let's start to integrate in $S$. From (\ref{pottot}) we straightforwardly get the result. It can be written in function of the various scales of the theory. Of course each expression makes sense  in the appropriate range of energy.:

\begin{eqnarray}
W(S)&=&S~\left(\ln\frac{\Lambda_I^{2N-N_f}~M^N~\prod_{f=1}^{N_f} m_f}{S^N}+N \right)\\
&=&S~\left(\ln\frac{\Lambda_{II}^{3N-N_f}\prod_{f=1}^{N_f} m_f}{S^N}+N \right)\label{potads}\\
&=&S~\left(\ln\frac{\Lambda_{III}^{3N}}{S^N}+N \right)
\label{potl}\end{eqnarray}
Consider then the superpotential (\ref{potl}). It has to be understood as the superpotential at energies much lower then the quarks' masses. As discussed, at those scales the theory we are considering reduces to pure ${\cal N}=1$ gluodynamics. It is immediate to see now that (\ref{potl}) is the standard Veneziano-Yankielowicz superpotential, whose minimization gives the correct value of the gaugino condensate.

We can now use (\ref{potads}) to integrate in the $N_f$ flavors and find thus the superpotential for both the gaugino condensate and the mesons in ${\cal N}=1$ supersymmetric QCD. We easily get:
\begin{equation}
W~=~\sum_im_iX_i+S~\left(\ln\frac{\Lambda_{II}^{3N-N_f}}{S^{N-N_f}\mbox{Det}X}+(N-N_f)\right)
\end{equation}
This is the Veneziano-Yankielowicz-Taylor superpotential\footnote{ Recent progresses towards a diagrammatic derivation of it have recently been made in \cite{Ambjorn:2003rp}.} \cite{Taylor:1982bp}.

Another interesting possibility we have is to use our starting point (\ref{pottotads}) to integrate in the chiral fundamental and anti-fundamental fields. The further relation we need is
\begin{equation}
\frac{\partial W_{low}}{\partial m_f}=X_f=\langle\tilde{Q}_{\tilde{f}}Q^f\rangle\end{equation} 
It's easy to get the result for one flavor. However, for generic number of flavors (always $N_f<N$) after having integrated in $q$ flavors we get the superpotential:
\begin{equation}
W=\sum_{i=1}^qm_iX_i+(N-q)(\prod_{i=1}^qX_i)^{\frac{1}{q-N}}(\eta^{2N}\Lambda_{II}^{3N-N_f})^{\frac{1}{N-q}}(\prod_{i=1}^{N_f-q}m_i)
\end{equation}
We can easily continue this procedure until we integrate all the $N_f$ flavors and we finally get:
\begin{equation}
W=\sum_{i=1}^{N_f}m_i X_i +(N-N_f)\left(\frac{\eta^{2N}\Lambda_{II}^{3N-N_f}}{\mbox{Det} X} \right)^{\frac{1}{N-N_f}}
\end{equation}
This is precisely the non-perturbative Affleck-Dine-Seiberg superpotential \cite{Affleck:1983mk}. Relations between this superpotential and matrix models were investigated in \cite{Argurio:2002xv,Demasure:2002sc,Demasure:2003sk}.

\section{On the vacua of ${\cal N}=1$ SYM theory}

Now we turn to consider more general superpotentials for the adjoint field but, for reasons explained in the introduction, we restrict ourselves to the case $N_f=0$. The aim is to study the structure of the moduli space of vacua. We deal with $U(N)$ gauge theory with ${\cal N}=1$ supersymmetry and a chiral superfield $\Phi$ in the adjoint representation. The superpotential is taken to be:
\begin{equation}
\label{suppot}
W(\Phi)~=~\sum_{k=0}^n\frac{g_k}{k+1}\mbox{Tr}~\Phi^{k+1}
\end{equation}

Our aim is to study the effective Lagrangian for the low energy gauge theory with gauge group $\prod_i U(N_i)$. A necessary assumption to do this is that the critical points of $W$ are distinct, so that all the components of $\Phi$ are massive classically. We then suppose that the underlying $U(N)$ gauge theory is weakly coupled at the scale set by those masses, so that the low energy physics is simply that of pure supersymmetric gauge theory with classical gauge group $\prod_i U(N_i)~\sim~U(1)^n\times\prod_{i=1}^nSU(N_i)$. The physical scale of this theory, $\Lambda_{II}$, is related to the ultraviolet one, $\Lambda_I$, by renormalization group matching conditions, namely:
\begin{equation}
M\Lambda_I^2~=~\Lambda_{II}^3.
\end{equation}
This theory has been recently extensively studied and it is known it has an effective superpotential \cite{Cachazo:2002ry}
\begin{equation}
\label{totsup}
W_{eff}=\sum_iN_i\frac{\partial {\cal F}_p(S_k)}{\partial S_i}+\frac{1}{2}\sum_{ij}\frac{\partial^2{\cal F}_p(S_k)}{\partial S_iS_j}w_{\alpha i}w_j^{\alpha}
\end{equation}
where $S_i$ are $U(N_i)$ composites of gauge fields 
\begin{equation}
S_i=\hat{S_i}-\frac{1}{2N_i}w_{\alpha i}w^{\alpha}_i\hspace{1cm},\hspace{1cm}w_{\alpha i}=\frac{1}{4\pi}\mbox{Tr}~W_{\alpha i}
\end{equation}
and $\hat{S}_i$ are $SU(N_i)$ composite fields. For small $S_i$, (following the notation of \cite{Cachazo:2002zk}) the superpotential (\ref{totsup}) reduces to:\begin{equation}
W_{\mbox{eff}}(S_i)=\sum_i\left(N_iS_i[\log(\Lambda_i^3/S_i)+1]+2\pi\mbox{i}\tau_0 S_i +2\pi {\mbox i}~b_iS_i\right)+....
\label{WLb}
\end{equation}
where $b_k=r_k-r_1$ are integer numbers and in these conventions $b_1=0$.

We can equivalently define this superpotential directly in terms of the $r_i$ variables, so that some aspects we are going to analyze will emerge more clearly. Then we get:
\begin{equation}
W_{\mbox{eff}}(S_i)=\sum_i\left(N_iS_i[\log(\Lambda_i^3/S_i)+1]+2\pi\mbox{i}\tau_0 S_i - 2\pi {\mbox i}~r_iS_i\right)+....
\label{WLr}
\end{equation}
where the integer $r_i$ are arbitrary.

The purpose of this section is to analyze the vacuum structure of such a theory, emphasizing the properties of the $\theta$-angle defining the different vacua. We will define an `effective' chiral symmetry useful in understanding this low energy physics. It will turn out that this is rather convenient in describing certain properties of the vacua with the same value of the confinement index (as it has been defined in \cite{Cachazo:2002zk}). In particular it will be possible to count them and find interpolation between vacua that are on different branches. Some of the properties of these vacua can be better understood referring to the multiplication map defined in \cite{Cachazo:2001jy}. We then briefly review it here elucidating its relations with the $\theta$-vacua.

\subsection{Multiplication map}

In \cite{Cachazo:2001jy} it was constructed a map that, for any positive integer $t$, associates vacua of the $U(N)$ theory with a given superpotential to vacua of the $U(tN)$ theory with the same superpotential. In particular in \cite{Cachazo:2002zk} it was shown that under this map, all the periods of the one form $T(z)~dz$, the generating function of the observables $\mbox{Tr}~\Phi^k$, are multiplied by $t$. It is thus easy to see that:
\begin{equation}
\theta_{tN}=t\theta_{N}\hspace{2cm},\hspace{2cm}\frac{1}{g_{tN}^2(\mu)}=\frac{t}{g_N^2(\mu)}
\label{mult}
\end{equation}
From this equation plus renormalization group considerations, it follows also that the dynamically generated scales of the two theories under consideration ($U(N)$ and $U(tN)$) are the same. These relations are consistent with what has been derived in \cite{Cachazo:2002zk}. Indeed also there the authors pointed out that:
\begin{equation}
\Lambda_N^{2N}=\eta^2\Lambda_{tN}^{2N}
\end{equation}
where $\eta^{2t}=1$. Here I would like to emphasize also that this relation (by holomorphy) naturally implies that $\theta_{tN}=t\theta_N+2\pi{\mbox i} k $ with $k=1,\ldots t$, consistently with ($\ref{mult}$).

These relations can also be generalized to the case of the low energy ${\cal N}=1$ SYM theory (whose physical scale is $\Lambda_{II}$). Indeed let us start considering the breaking $U(N)\to U(1)\times SU(N)$. Following \cite{Cachazo:2002zk} and using Chebyshev polynomial (see Appendix A), we see that this occurs when the factorization of SW curve occurs with the following function:
\begin{equation}
F_2(x)=x^2-4\rho^2\Lambda_{I,N}^2
\end{equation}
where $\rho^{2N}=1$. From this, using the relations
\begin{equation}
F_2=\frac{1}{M^2}(W'(x)^2+f_0(x))\hspace{1cm},\hspace{1cm}S=-\frac{1}{4M}f_0
\end{equation}
we easily find
\begin{equation}
S=\rho^2M\Lambda_{I,N}^2=\rho^2\Lambda_{II,N}^3
\end{equation}
Now it is easy to see that under multiplication map, we have
\begin{equation}
F_2(x)=x^2-4\rho^2\eta^{-2/N}\Lambda_{tN}^2
\end{equation}
where again $\rho^{2N}=1$ and $\eta^{2t}=1$. Then we find the following relations for the phases of the gaugino condensate ($\varphi$) in the two theories, (let me add now an extra index `i' to generalize these considerations to the case of $U(N)\to \prod_i U(N_i)$):
\begin{equation}
\varphi_{N_i}=\frac{\theta_0}{N_i}+\frac{2\pi r_i}{N_i}\hspace{1,5cm},\hspace{1,5cm}\varphi_{tN_i}=\frac{\theta_0}{N_i}+\frac{2\pi t r_i}{t N_i}-\frac{2\pi l_i}{t N_i}\label{phis}
\end{equation}
with $r_i=1,\ldots,N_i$ and $l_i=0,\ldots t-1$. From this expression the relations between the $\theta$-angles of the two related theories follow in a straightforward way. Indeed, for each value of the angle of the $U(N_i)$ theory (equivalently for fixed $r_i$), there are $t$ corresponding angles in $U(tN)$ (corresponding to $l_i=0,\ldots,t-1$), namely (ignoring the dependence on ``i''):
\begin{equation}
\theta_{t N}=2\pi(r t - l)+t\theta^0_N
\label{multt}\end{equation}
In the spirit of \cite{Cachazo:2002zk} we can see that this $l$ is the parameter related to the quantum numbers of the condensing object in the confining $SU(N_i)$ theory, (namely $W^lH$, as defined in \cite{Cachazo:2002zk}, do condense).

From \cite{Cachazo:2002zk} we know also that in a branch of confining index $t$ we have at least $t$ vacua. As we are going to count them, let's review here that argument. It relies on the multiplication map just defined. We start from a Coulomb vacuum in the theory $U(N)\to\prod U(N_i)$ to study the $t$ confining vacua we have in $U(tN)\to \prod U(tN_i)$. The $b_i=r_{i+1}-r_i$ label different $SU(N_i)$ vacua. We are considering a Coulomb vacuum in $U(N)$. Then we have that the greatest common divisor between the $N_i$ and the $b_i$, (that we label as $(N_i,b_i)$), is 1. Under the multiplication map, it follows that $(t N_i,\tilde{b}_i)=t$, where with the tilde we refer to the $U(tN)$ quantities ($\tilde{b}_i=t b_i$). Now let's consider $\tilde{r}_i=u_i+tr_i$ with $u=0,\ldots,t-1$. We have then that now $\tilde{b}_i=(u_{i+1}-u_i)+t(r_{i+1}-r_i)$. It is easy to see that if $u_i=u_{i+1}$ we have that also now $(tN_i,\tilde{b}_i)=t(N_i,b_i)=t$ and we have then found the $t$ vacua with confinement index $t$. It follows also that if $u_i\neq u_{i+1}$, $(tN_i,\tilde{b}_i)\neq t$. Then we recover the result that for every Coulomb vacuum of $U(N)$ we find precisely $t$ confining vacua with index $t$ in $U(tN)$ and that we can have no more.

In the next subsection, using arguments related to the symmetries of the low energy effective action, we will count these vacua in a different way, finding then a different (but not contradictory) result.

\subsection{Global symmetries}

Let us come to the analysis of the global symmetries of such a theory. In terms of the microscopic degrees of freedom, it has two continuous global $U(1)$ symmetries \cite{Cachazo:2002ry}:

\begin{tabular}{ccccc} $\ $  & $\phi$ & $W_{\alpha}$ & $g_l$ & $\Lambda^{2N}$ \\
$Q_{\phi}$ & 1 & 0 & $-(l-1)$ & $2N$ \\
 $Q_{\theta}$ & 0 & 1 & 2 & 0 \end{tabular}\\ 
These are the standard symmetries of the adjoint theory. But let us consider an `effective' symmetry that is present at low energy, when the adjoint scalar is integrated out and we are left with just ${\cal N}=1$ SYM theory with gauge group $\prod_{i=1}^n SU(N_i)$ and some trivial free abelian theory. As long as the couplings between the different gauge factors are negligible, the different gauge groups don't interact and we can then define $n$ such effective symmetries, that act in the following way:

\begin{tabular}{ccc} $\ $ &  $W_{\alpha}^i$ & $\Lambda_i^{3N_i}$\\ $Q_{R_j}$ &  $\delta^i_j$ & $2N_i\delta_i^j$ \end{tabular}\\ 
Out of these symmetries it is possible to define a diagonal one (let's call it $U(1)_R$) that satisfies two natural requirements:
\begin{itemize}\item all the $W_{\alpha}^i$ have integer charge,
\item the (unique) well defined scale of the low energy theory ($\Lambda_{II}^{3N}=\Lambda_I^{2N} g_1^N$) has definite charge under this symmetry.
\end{itemize}
We then get the following charges for this symmetry: 

\begin{tabular}{ccc} $\ $ &  $W_{\alpha}^i$ & $\Lambda_i^{3N_i}$\\ $Q_{R}$ & $\frac{m}{N_i}$ & $2m$ \end{tabular}\\ 
where $m$ is the minimum commune multiple of the $N_i$ (that we indicate with $[N_i]$).

It is now easy to see that under this symmetry all the $\theta_i$ angles are shifted by the same quantity, namely $\theta_i\to\theta_i+2m\alpha$ when $W_{\alpha}^i\to\exp(\frac{i m \alpha}{N_i})W_{\alpha}^i$ (such that the composite field $\hat{S}_i\to\exp(\frac{2i m \alpha}{N_i})\hat{S}_i$). This would correspond to shift all the $r_i$ appearing in the Lagrangian (\ref{WLr}) by $2m\alpha$. Moreover, from the form of the Lagrangian (\ref{WLr}), we notice that this corresponds to the action of a global $U(1)$ symmetry under which the $\theta$ angle of the underlying $U(N)$ theory shifts by $2m\alpha$, leaving the $r_i$ unchanged. We can thus also see that $\Lambda_{II}^{3N}$ has charge 2m as the $\Lambda_i^{3N_i}$. Indeed this can be verified by threshold matching conditions that relates the different physical scales: 
\begin{equation}
\Lambda_{II}^{3N}\sim\Lambda_i^{3N_i}\Sigma_i\end{equation}
where the $\Sigma_i$ are dimensional factor neutral under $U(1)_R$.

It follows now that only a discrete subgroup of this $U(1)_R$ is really a symmetry for the low energy effective theory. Indeed from the action of this symmetry on the $\theta$-angle we have that only discrete values of $\alpha$ are allowed, namely
\begin{equation}
\alpha_k~=~\frac{\pi k}{m}
\label{k}
\end{equation}
Now, from the charges of the superfields $W_{\alpha}^i$ it is clear that the action of the symmetry reduces to the identity operation when $\alpha=2 l\pi$, with $l$ an integer. From this and (\ref{k}) it immediately follows that the actual `effective' symmetry is a $Z_{2m}$ subgroup of $U(1)_R$. Moreover, in each vacuum, fixed by the values of the $S_i$, this symmetry is spontaneously broken to $Z_2$. We have then $m$ distinct vacua, permuted by this $Z_{2m}$ symmetry. All these vacua are contained in the Lagrangian (\ref{WLr}), with fixed values of the parameters $r_i$ but for different values of the $\theta$ angle.
However these different $\theta$-vacua are related by shifts of integer multiples of $2\pi$.
On physical grounds, it appears then natural that all these vacua have the same value of the confinement index. Indeed they are related by a symmetry and we expect the physics to be $2\pi$ periodic in the $\theta$ angle of the underlying $U(N)$ theory. 

Then we claim rather generally that, given a partition of $N_i$ and  a partition of $b_i$ such that the confinement index is $t$, there are $m$ (or a multiple of $m$) vacua with $t$ as confinement index. These vacua are naturally permuted by the discrete chiral $Z_{2m}$ low-energy symmetry. 

The mathematical relation which proves this statement is the following number theory property\footnote{This is related to the fact that the $r_i$ are only defined modulo $N_i$}:
\begin{eqnarray}
(N_1,N_2,\ldots,N_n,b_2,b_3,\ldots,b_{n})=\\ \nonumber(N_1,N_2,\ldots,N_n,b_2+\alpha_2N_2+\beta_2N_1,b_3+\alpha_3N_3+\beta_3N_1,\ldots,b_{n}+\alpha_nN_n+\beta_nN_1)
\end{eqnarray}
with $\alpha_I$ and $\beta_I$ arbitrary integers.

A non trivial consequence that follows from this is that there are always a multiple of $m$ vacua for every value of the confinement index. It is moreover worthwhile to note that these $m$ vacua, connected by the $Z_{2m}$ effective symmetry, can in general be on different branches\footnote{These branches were defined in \cite{Cachazo:2002zk}.} of the parameter space. 
This discrete chiral symmetry can then, at least partially, motivate the presence of many branches. 

To understand how this is related to the properties of the multiplication map it is better to see how it works in a simple example, that we work out in appendix B.

\section{Conclusions}

In this paper I have determined the value of the gaugino condensate in ${\cal N}=1$ $SU(N)$ gauge theory. My findings are in agreement with the literature. In particular they reproduce the result of the `weak coupling' approach.

The structure of the parameter space of the theory is analyzed, finding relations between vacua with the same value of the confinement index.

All this analysis has been performed using purely field theoretic tools. However there are many analogies with some recent string theory findings. The embedding of this field theoretic approach in string/M theory surely deserves investigation.

\section{Acknowledgments}

I am grateful to Yves Demasure, Paolo DiVecchia and Shigeki Sugimoto for very useful discussions.

This work has been supported by the European Community's Human Potential
Programme under contract HPRN-CT-2000-00131 Quantum Spacetime.

\appendix
 
\section{Chebyshev polynomials}

There are two kind of Chebyshev polynomials (${\cal T}_N(z)$ and ${\cal U}_{N-1}(z)$). They are defined as follows:
\begin{equation}
{\cal T}_N(z)~=~\cos(N\theta)\hspace{3cm}{\cal U}_{N-1}(z)~=~\frac{1}{N}\frac{d{\cal T}_N}{dz}(z)~=~\frac{\sin(N\theta)}{\sin\theta},
\end{equation}
where $z=\cos\theta$, being thus both $\cos(N\theta)$ and $\sin(N\theta)/\sin\theta$ polynomials in $z$. They are related by an identity that can be easily checked:
\begin{equation}
{\cal T}_N(z)^2-1~=~(z^2-1)~{\cal U}_{N-1}(z)^2.
\end{equation}
 
\section{Example}

Let us consider now the theory $U(5)\to U(2)\times U(3)$. At low energy, in the semiclassical limit, it is possible to see the six vacua of this theory as the product of two vacua $a$ of $U(2)$, (with $a=1,2$ labeling for example the values of the phase of $S_1$, related by $2\pi$ shift in $\theta_2$), with three vacua $a'$ of $U(3)$, (analogously with $a'=1,2,3$). Then the six vacua are the following:

\begin{tabular}{ccccc} $a'$  & $a$ & $\theta_5$ & $b$ & $t$ \\
1 & 1 & 0& 0 & 1\\
2& 2& $2\pi$ & 0 & 1\\
3& 1& $4\pi$ &2 & 1\\
1& 2& $6\pi$ &1 & 1\\
2& 1& $8\pi$ &1 & 1\\
3 & 2& $10\pi$& 1 & 1
\end{tabular}\\ 
All these vacua are Coulomb vacua and these are all the vacua for this choice of $N_i$. It's worthwhile to stress that all these vacua are vacua of the same low energy effective Lagrangian (\ref{WLr}) with the choice $r_1=r_2$. This is immediately clear if we realize that the $r_i$ parameter are defined only modulo $N_i$. Notice that, accordingly to what stated in the preceding subsection, here we precisely have 6 vacua ($[3,2]=6$) with the same value of the confinement index ($t=1$). 

Now let's consider $U(10)\to U(4)\times U(6)$. Some of the vacua of this theory should then be obtained by multiplication map from the vacua of the $U(5)$ theory just described. For the time being, with analogous notation, we build all the $U(10)$  vacua:

\begin{tabular}{ccccccccccc} $a'$  & $a$ & $\theta_{10}$ & $b$ & $t$ & $\hspace{2cm} | \hspace{2cm}$ &  $a'$  & $a$ & $\theta_{10}$ & $b$ & $t$ \\
1 & 1 & 0 & 0&2   &$\hspace{2cm} | \hspace{2cm}$ &1&2&0&1&1\\
2& 2& $2\pi$ & 0&2 &$\hspace{2cm} | \hspace{2cm}$ &2&3&$2\pi$&1&1\\
3& 3& $4\pi$ &0&2 &$\hspace{2cm} | \hspace{2cm}$ &3&4&$4\pi$&1&1\\
4& 4& $6\pi$ &0&2 &$\hspace{2cm} | \hspace{2cm}$ &4&5&$6\pi$&1&1\\
1& 5& $8\pi$ &4&2 &$\hspace{2cm} | \hspace{2cm}$ &1&6&$8\pi$&5&1\\
2 & 6& $10\pi$& 4&2 &$\hspace{2cm} | \hspace{2cm}$ &2&1&$10\pi$&1&1\\
3&1 &$12\pi$& 2&2 &$\hspace{2cm} | \hspace{2cm}$ &3&2&$12\pi$&1&1\\
4&2&$14\pi$&2&2 &$\hspace{2cm} | \hspace{2cm}$ &4&3&$14\pi$&1&1\\
1&3&$16\pi$&2&2 &$\hspace{2cm} | \hspace{2cm}$ &1&4&$16\pi$&3&1\\
2&4&$18\pi$&2&2 &$\hspace{2cm} | \hspace{2cm}$ &2&5&$18\pi$&3&1\\
3&5&$20\pi$&2&2 &$\hspace{2cm} | \hspace{2cm}$ &3&6&$20\pi$&3&1\\
4&6&$22\pi$&2&2 &$\hspace{2cm} | \hspace{2cm}$ &4&1&$22\pi$&3&1
\end{tabular}

These are all the vacua for this choice of $N_i$.
It is now easy to see how the $t=2$ vacua of $U(10)$ (left part of the table) are related to the Coulomb vacua of the $U(5)$ table. For example, the first is related to the $\theta_5=0$ vacuum, (see (\ref{multt}) with $r=1,\ t=2,\ l=2$). And so on $\ldots$
Consider now as all these $t=2$ vacuum are related by the $Z_{2m}$ symmetry we discussed. Actually they are related by a subgroup of it ($Z_{m}$), analogously to what happen in standard ${\cal N}=1$ $SU(N)$ gluodynamics. Indeed they are precisely $m$ (in the case at hand $m=[4,6]=12$). This $Z_{12}$ symmetry (more generally $Z_{m}$) relates also the vacua of the second column, with $t=1$. Finally let us stress that all the $t=2$ vacua are described by the same low energy effective Lagrangian (\ref{WLr}) with $r_1=r_2$, while the $t=1$ are described again by (\ref{WLr}), but now with $r_2=r_1+1$.

\end{document}